# Counting Skolem Sequences

Ali Assarpour     Amotz Bar-Noy     Ou Liu

May 5, 2017

**Abstract**

We compute the number of solutions to the Skolem pairings problem, $S(n)$, and to the Langford variant of the problem, $L(n)$. These numbers correspond to the sequences $A059106$, and $A014552$ in Sloane's Online Encyclopedia of Integer Sequences. The exact value of these numbers were known for any positive integer $n < 24$ for the first sequence and for any positive integer $n < 27$ for the second sequence. Our first contribution is computing the exact number of solutions for both sequences for any $n < 30$. Particularly, we report that

$S(24) = 102, 388, 058, 845, 620, 672.$

$S(25) = 1, 317, 281, 759, 888, 482, 688.$

$S(28) = 3, 532, 373, 626, 038, 214, 732, 032.$

$S(29) = 52, 717, 585, 747, 603, 598, 276, 736.$

$L(27) = 111, 683, 611, 098, 764, 903, 232.$

$L(28) = 1, 607, 383, 260, 609, 382, 393, 152.$

Next we present a parallel tempering algorithm for approximately counting the number of pairings. We show that the error is less than one percent for known exact numbers, and obtain approximate values for $S(32) \approx 2.2 \times 10^{26}$, $S(33) \approx 3.6 \times 10^{27}$, $L(31) \approx 5.3 \times 10^{24}$, and $L(32) \approx 8.8 \times 10^{25}$.

## 1 Introduction

### 1.1 Background

In 1957, TH. Skolem, [22, 23], a Norwegian mathematician, considered the problem of distributing the numbers $1, 2, 3, \ldots, 2n$ in $n$ pairs $(a_r, b_r)$ such that $b_r - a_r = r$ for $r = 1, 2, 3, \ldots, n$. For example when $n = 4$, one possible distribution of the numbers $1, 2, 3, 4, 5, 6, 7, 8$ in four pairs with differences $1, 2, 3, 4$ is $(7, 8)\ (3, 5)\ (1, 4)\ (2, 6)$. He named such pairings a '1, +1 system'.

At about the same time, C.D. Langford, [11], a Scottish mathematician, devised a similar problem while observing his son playing with a set of colored blocks. He noticed that his son had arranged a set of six blocks, two red, two blue, and two green, such that the pair of red blocks were separated by a single block, the pair of blue blocks were separated by two blocks, and the pair of green blocks were separated by three blocks. Furthermore, he noticed he can add a pair of yellow blocks to the arrangement in a way that would preserve the distances of the previous blocks while having the yellow pair be separated by four blocks.

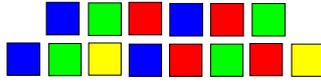

Capturing this idea using numbers, he asked the following question: *"given a sequence of 2n numbers $\{1,1,2,2,\ldots,n,n\}$, is there a permutation in which the two copies of each number k are k units apart?"* For instance, when $n = 4$ the sequence 2, 3, 4, 2, 1, 3, 1, 4 is one such permutation.

In 1966, Nickerson [16], proposed a variant of Langford's problem in which the pair of numbers $k$ should be separated by $k - 1$ (instead of $k$) other numbers. This variant is equivalent to Skolem's '1, +1 system'. For instance, when $n = 5$ the sequence 3,5,2,3,2,4,5,1,1,4 is equivalent to partitioning the numbers $1, \ldots, 10$ into the five pairs $(8, 9)$ $(3, 5)$ $(1, 4)$ $(6, 10)$ $(2, 7)$ with the differences 1, 2, 3, 4, and 5.

| 1 | 2 | 3 | 4 | 5 | 6 | 7 | 8 | 9 | 10 |
|---|---|---|---|---|---|---|---|---|----|
| 3 | 5 | 2 | 3 | 2 | 4 | 5 | 1 | 1 | 4  |

The first natural question to ask is: *"for which n do such pairings exist?"* That is, for a given $n$, we are looking for necessary and sufficient conditions for the existence of such a sequence. In the case of the Skolem problem the existence question was answered by Skolem in 1957. He showed that a pairing exists iff $n \equiv 0, 1 \mod 4$. His result follows the observation that all the $n$ pairs occupy the positions 1 to $2n$ in some order and that the first occurrence of $r$ is at position $a_r$ while the second is at position $a_r + r$. A method for constructing such sequences was presented as well. It is reproduced in table 1 below.

| | |
|---|---|
| $n = 4k$ | $(4k + r, 8k - r)$ $r = 0, \ldots, 2k - 1$, <br> $(2k + 1, 6k)$, $(2k, 4k - 1)$, <br> $(r, 4k - 1 - r)$ $r = 1, \ldots, k - 1$, <br> $(k, k + 1)$, <br> $(k + 2 + r, 3k - 1 - r)$ $0, \ldots, k - 3$. |
| $n = 4k + 1$ | $(4k + 2 + r, 8k + 2 - r)$ $r = 0, \ldots, 2k - 1$, <br> $(2k + 1, 6k + 2)$, $(2k + 2, 4k + 1)$, <br> $(r, 4k + 1 - r)$ $r = 1, \ldots, k$, <br> $(k + 1, k + 2)$, <br> $(k + 2 + r, 3k + 1 - r)$ $r = 1, \ldots, k - 2$. |

Table 1: Skolem's construction of a solution to the Skolem pairing. The graphical representation is for $n = 8$. When $|b_r - a_r| = r$ is even the pair is connected by a blue line, otherwise the pair is connected by a red line

In 1959, Davies [6] answered the existence question for the Langford formulation. He showed that a sequence exists iff $n \equiv 0, 3 \mod 4$ using similar arguments to those using to show the existence of the Skolem's sequences. He also provided a method for constructing such a sequence. This method is reproduce in table 2 below.

| | | |
|---|---|---|
| $n = 4k-1$ | 4k-4, ..., 2k, 4k-2, 2k-3, ..., 1, 4k-1, 1, ..., 2k-3, 2k, ..., 4k-4, 2k-1, 4k-3, ..., 2k+1, 4k-2, 2k-2, ..., 2, 2k-1, 4k-1, 2, ..., 2k-2, 2k+1, ..., 4k-3 | |
| $n = 4k$ | 4k-4, ..., 2k, 4k-2, 2k-3, ..., 1, 4k-1, 1, ..., 2k-3, 2k, ..., 4k-4, 4k, 4k-3, ..., 2k+1, 4k-2, 2k-2, ..., 2, 2k-1, 4k-1, 2, ..., 2k-2, 2k+1, ..., 4k-3, 2k-1, 4k | |

Table 2: Davies' construction of a solution to the Langford pairing. The graphical representation is for $n = 8$ where the even pairs are connected by a blue lines and the odd pairs by red lines

## 1.2 Related Work

The first generalization, called *hooked Skolem sequence*, was introduced by Skolem [23]. Here the numbers $1, 2, \ldots, 2n-1, 2n+1$ are distributed into $n$ pairs with differences $1, 2, \ldots, n$. Such sequences were further studied by O'Keefe [18]. A variation of the hooked sequence was studied by Rosa [20] were the skip was allowed to occur in the middle of the sequence. Later Abrham and Kotzig [2] generalized such sequences by allowing the skip to occur anywhere in the sequence. They called these sequences the *extended Skolem sequences*. Further generalizations, called the *near Skolem sequence* and the *near hooked Skolem sequence*, were later introduced by Stanton and Goulden [25] in 1981, In 1991, another generalization, called the *disjoint Skolem sequence*, was introduced by Shalaby [21]. The adaptation of sequences to graphs, called *Skolem labeling*, was introduced by Mendelsohn and Shalaby [14], and was studied in paths, trees and cycles. Finally, in 2007, Nordh [17] studied Skolem sequences with different set of integers and set of differences and introduced the *perfect Skolem set*, *multi Skolem set*, and the *generalized multi Skolem set* problems. Many other generalizations have been studied as well due in part to their importance in design theory.

## 1.3 Contributions

In this note we are concerned with a second question that is frequently asked: *"how many sequences are there?"* There are two counts associated with the Skolem sequence in Sloane's Online Encyclopedia of Integer Sequences (OEIS) [24] $A004075$ and $A059106$. The former counts all the sequences while the later considers reflected sequences to be equivalent. For instance, the pair of sequences $4, 2, 3, 2, 4, 3, 1, 1$ and $1, 1, 3, 4, 2, 3, 2, 4$ are considered to be distinct sequences in the first and the same in the second. When $n > 1$, the count of these sequences are related to each other by a factor of two, because no sequence is a palindrome. Similarly, there are two counts for the Langford sequence in the OEIS, $A176127$ and $A014552$, where the second counts reflected sequences as the same. John Miller maintains a dedicated web page [15] which documents many aspects of these sequences. Table 3 is taken from his website and summarizes the finding of the known counts of these sequences.

| Problem | Date | Person | Computer | Time | Language |
|---|---|---|---|---|---|
| $S(4-5)$ | Feb-69 | John Miller | IBM 1130 | ? | ? |
| $S(8-9)$ | Feb-69 | John Miller | IBM 1130 | ? | ? |
| $S(12-13)$ | Mar-89 | John Miller | VAX | ? | ? |
| $S(16)$ | Feb-99 | John Boyer | Intel | ? | ? |
| $S(17)$ | Feb-99 | John Boyer | Intel | ? | ? |
| $S(20)$ | Mar-02 | Mike Godfrey | Pentium III | 65.5 hours | FORTRAN |
| $S(21)$ | Mar-02 | Godfrey/van Bruchem | AMD/Pentium | <week | FORTRAN |
| $L(3-4)$ | ? | C. Dudley Langford | Hand | ? | ? |
| $L(7)$ | 1951 | Roy O. Davies | Hand | ? | ? |
| $L(7-8)$ | May-67 | Dave Moore | TRW-130 | $5m, 40m$ | FORTRAN |
| $L(7-8)$ | Nov-67 | Glen F. Stahly | ? | ? | ? |
| $L(7-8)$ | Nov-67 | John Miller | IBM 1130 | ? | FORTRAN |
| $L(7-8)$ | Nov-67 | Malcolm Holtje | ? | ? | ? |
| $L(7-8)$ | Nov-67 | Robert Smith | ? | ? | ? |
| $L(7)$ | Nov-67 | Thomas Starbird | ? | ? | ? |
| $L(7-12)$ | Nov-67 | E. J. Groth | SDS 930 | <day | FORTRAN |
| $L(11-12)$ | 1968? | John Miller | IBM 1130 | ? | Asm |
| $L(15)$ | Sep-80 | John Miller | VAX 11/780 | ? | Pascal |
| $L(15)$ | Feb-87 | Frederick Groth | Commodore 64 | 15.5 days | Asm |
| $L(16)$ | Feb-87 | Frederick Groth | Commodore 64 | 122.4 days | Asm |
| $L(15)$ | Jul-89 | Andrew Burke | Cogent XTM | ? | C |
| $L(16)$ | Jul-89 | Andrew Burke | Cogent XTM | 120hours | C |
| $L(16)$ | May-94 | John Miller | Dec Alpha | ? | ? |
| $L(19)$ | May-99 | Rick Groth Team | Mac/Pentium | 2 months | C |
| $L(19)$ | Jul-99 | John Miller | DEC Alpha | 2.5 years | C |
| $L(19)$ | Mar-02 | Ron van Bruchem | Pentium | 6 hours | FORTRAN |
| $L(20)$ | Feb-02 | Godfrey/van Bruchem | AMD/Pentium | 1 week | FORTRAN |
| $L(23)$ | Apr-04 | Krajcki Team | Sun/Intel | 4 days | Java/CONFIIT |
| $L(24)$ | Apr-05 | Krajcki Team | $12-15$ processors | 3 months | Java/CONFIIT |

Table 3: Contributors to the count of the number of Skolem and Langford sequences, from Miller's web page[15]

Our first contribution extends the count for the Skolem and Langford sequences by finding the exact values for the previously unknown $S(24)$, $S(25)$, $S(28)$, $S(29)$, $L(27)$ and $L(28)$. These values were computed using an Algebraic counting technique first proposed by Godfrey,[15, 12, 7], in 2002 for counting the number of Langford sequences. The algorithm was implemented and executed on NVIDIA's CUDA parallel computing platform. The unprecedented amount of parallelism available in the Graphics Processing Units makes it a natural platform to accelerate combinatorial counting problems. However they are not an ideal platform as there can be "silent errors" [27] which may lead to incorrect output. In order to catch any silent errors and reduce the possibility of an erroneous result each GPU job was ran twice on separate GPUs. Incase an error was presented and the result did not match, the job was ran a third time to establish a match and determine the correct result. The newly found Skolem and Langford counts are shown in tables 4 and 5 respectively.

| $n$ | $A059106$ | NVIDIA Kepler | Time |
|---|---|---|---|
| 24 | $102,388,058,845,620,672$ | 1 GPU | < day |
| 25 | $1,317,281,759,888,482,688$ | 4 GPU's | < day |
| 28 | $3,532,373,626,038,214,732,032$ | ~ 32 GPU's | ~ 9 days |
| 29 | $52,717,585,747,603,598,276,736$ | ~ 32 GPU's | ~ 8 weeks |

| $n$ | $A004075$ |
|---|---|
| 24 | $204,776,117,691,241,344$ |
| 25 | $2,634,563,519,776,965,376$ |
| 28 | $7,064,747,252,076,429,464,064$ |
| 29 | $105,435,171,495,207,196,553,472$ |

Table 4: Newly computed values for the Skolem sequence.

In our second contribution, we propose a method to approximately count these sequences. To the best of our knowledge this is the first attempt to provide non-exact values. The procedure is a Markov based algorithm. Experimentally, we showed that there was no error for $n < 12$ and that the error was less than one percent for $12 \leq n \leq 29$. Using this method we report the following approximate values: $S(32) \approx 2.2 \times 10^{26}$, $S(33) \approx 3.6 \times 10^{27}$, $L(31) \approx 5.3 \times 10^{24}$, and $L(32) \approx 8.8 \times 10^{25}$.

| $n$ | $A014552$ | NVIDIA Kepler | Time |
|---|---|---|---|
| 27 | $111,683,611,098,764,903,232$ | $\sim 32$ GPU's | $\sim 2$ days |
| 28 | $1,607,383,260,609,382,393,152$ | $\sim 32$ GPU's | $\sim 9$ days |

| $n$ | $A176127$ |
|---|---|
| 27 | $223,367,222,197,529,806,464$ |
| 28 | $3,214,766,521,218,764,786,304$ |

Table 5: Newly computed values for the Langford sequence. $L(27)$ and $L(28)$ have since been confirmed as reported by Miller [15]

## 2 Exact Counting Algorithm

There are three approaches to count the number of sequences. The first approach, proposed by Miller [15], is to systematically generate all possible valid pairings and count them. The algorithm proceeds by placing the pairs in decreasing order, starting with leftmost available position where the pair can fit into. Once the pair is placed the algorithm tries to place the next smaller pair, if it can not be placed then the previously placed pair needs to be moved to the next available valid position. The algorithm stops when all possible positions for the largest pair have been explored. A second approach first proposed by Godfrey, [15, 12, 7], in 2002, is to model the problem by a polynomial where each term represents the label and its position, $F(n, X) = \prod_{i=1}^{n} \sum_{k=1}^{2n-i} x_k x_{k+i}$. The number of pairings is then the coefficient of the term $x_1 x_2 \ldots x_{2n}$. A third approach is based on the principle of inclusion exclusion, first proposed by Larsen, [12], in 2009. Letting each set, $A_i$, be the set of invalid pairings that avoid the position $i$ in the universe of all possible pairings, $A_\emptyset$. The valid pairings then are those that avoid no positions, $\overline{A_1} \cap \ldots \cap \overline{A_{2n}}$. The cardinality of this set is the number of sequences, $|A_\emptyset| - |A_1 \cup \ldots \cup A_{2n}|$.

We implement Godfrey's algebraic method on NVIDIA's CUDA parallel computing platform to count the number of sequence. We present the details of the algorithms in the following section from the perspective of counting Skolem sequences, while considering reflected sequences as distinct. The algorithm is easily modifiable to count the Langford sequences instead.

### 2.1 Algebraic Counting Method

An algebraic method for counting the number of sequences was proposed by Godfrey in 2002. There is no official paper on the algorithm but the description of this method appears in [7] and [15]. In this approach the problem is modeled by a polynomial where each term represents the label and its position, and the number of pairings is then the coefficient of the term $x_1 x_2 \ldots x_{2n}$.

$$F(n, X) = \prod_{i=1}^{n} \sum_{k=1}^{2n-i} x_k x_{k+i}$$

For instance when $n = 4$, $X = (x_1, x_2, x_3, x_4, x_5, x_6, x_7, x_8)$, and

$$\begin{aligned} F(4, X) =& (x_1 x_2 + x_2 x_3 + x_3 x_4 + x_4 x_5 + x_5 x_6 + x_6 x_7 + x_7 x_8) \\ & (x_1 x_3 + x_2 x_4 + x_3 x_5 + x_4 x_6 + x_5 x_7 + x_6 x_8) \\ & (x_1 x_4 + x_2 x_5 + x_3 x_6 + x_4 x_7 + x_5 x_8) \\ & (x_1 x_5 + x_2 x_6 + x_3 x_7 + x_4 x_8) \end{aligned}$$

where each of the factors represents the possible ways in which a label and position can appear in the solution. For instance the pair of $4's$ can be placed in the first and fifth or second and sixth or third and seventh or fourth and eight positions, represented by the factor $(x_1 x_5 + x_2 x_6 + x_3 x_7 + x_4 x_8)$. When the polynomial is expanded, the coefficient of the term where all variables appear is then the number of possible pairings (twice that if reflected solutions are considered the same). When $F(4, X)$ is expanded the coefficient of the the term $x_1 x_2 x_3 x_4 x_5 x_6 x_7 x_8$ is 6, which correspond to sequences

| | |
|---|---|
| 4, 2, 3, 2, 4, 3, 1, 1 | $(x_7 x_8)(x_2 x_4)(x_3 x_6)(x_1 x_5)$ |
| 3, 4, 2, 3, 2, 4, 1, 1 | $(x_7 x_8)(x_3 x_5)(x_1 x_4)(x_2 x_6)$ |
| 4, 1, 1, 3, 4, 2, 3, 2 | $(x_2 x_3)(x_6 x_8)(x_4 x_7)(x_1 x_5)$ |
| 2, 3, 2, 4, 3, 1, 1, 4 | $(x_6 x_7)(x_1 x_3)(x_2 x_5)(x_4 x_8)$ |
| 1, 1, 4, 2, 3, 2, 4, 3 | $(x_1 x_2)(x_4 x_6)(x_5 x_8)(x_3 x_7)$ |
| 1, 1, 3, 4, 2, 3, 2, 4 | $(x_1 x_2)(x_5 x_7)(x_3 x_6)(x_4 x_8)$ |

Computing this coefficient by expanding the polynomial is rather difficult and just as time consuming as generating and counting the number of solutions since there are $\frac{(2n-1)!}{(n-1)!}$ terms. However evaluating a polynomial is relatively easy. Therefore, to obtain the relevant coefficient, Godfrey suggests evaluating the polynomial while allowing each variable to take on the values $1$ and $-1$, and summing resulting value of the product of the variables with valuation of the polynomial over all possible choices for the variables,

$$\sum_{(x_1, \ldots, x_{2n}) \in \{1, -1\}} \left( \prod_{i=1}^{2n} x_i \right) F(n, X)$$

**Lemma 2.1.**

$$S(n) = \left( \frac{1}{2^{2n}} \right) \sum_{(x_1, \ldots, x_{2n}) \in \{1, -1\}} \left( \prod_{i=1}^{2n} x_i \right) \prod_{i=1}^{n} \sum_{k=1}^{2n-i} x_k x_{k+i}$$

*Proof.* To see the result, first notice that there are $2^{2n}$ evaluations of the polynomial. Next consider each term in $F(n, X)$ other than $x_1 \ldots x_{2n}$, each of them is missing at least one variable. Consider the case when exactly one variable is missing, say $x_i$, then these terms when multiplied by $x_i$ and summed over choice of $x_i = +1$ and $x_i = -1$ result in zero. Now consider what happens when there are $k$ missing variables, $x_{i_1}, \ldots, x_{i_k}$, then there are $2^k$ choices for settings these variables, exactly half of the time the product of $x_{i_1} \ldots x_{i_k}$ evaluates to $+1$ and the other half the product evaluates to $-1$. For a fixed choice for all other variables, summing over all choices for $x_{i_1} \ldots x_{i_k}$ results in zero. At the end what remains is then just $2^{2n}$ times the coefficient of the term $x_1 \ldots x_{2n}$. Finally the value of $S(n)$ is found simply by dividing the result by $2^{2n}$. □

---

**Algorithm 1:** Godfrey's algorithm for counting the number of Skolem sequences of order $N$

**Data:** $N$ number of pairs to be placed
**Result:** *count* number of valid sequences
**Algorithm** Godfrey()
    $S \leftarrow 0$;
    **for** $X \in \{+1, -1\}^{2n}$ **do**
        $signX \leftarrow 1$;
        **for** $i \leftarrow 1$ **to** $2n$ **do**
            $signX \leftarrow signX \times X(i)$;
        **end**
        $prod \leftarrow 1$;
        **for** $i \leftarrow 1$ **to** $n$ **do**
            $sum \leftarrow 0$;
            **for** $k \leftarrow 1$ **to** $2n - i$ **do**
                $sum \leftarrow sum + X(k)X(k+i)$;
            **end**
            $prod \leftarrow prod \times sum$;
        **end**
        $S \leftarrow S + signX \times prod$;
    **end**
    **return** $\frac{S}{2^{2n}}$;

---

A simple implementation of this procedure makes $2^{2n}$ evaluations of the polynomial, where each evaluation costs $O(n^2)$ multiplications and additions, for a total cost of $O(n^2 2^{2n})$. The implementation can be improved considerably by taking advantage of Gray code ordering of all the possible settings of $X$. By using such an ordering the evaluation cost of the polynomial is reduced to a linear number additions and multiplications when going from one setting to the next. Reducing the running time to $O(n2^{2n})$. Further speed ups are achieved by using symmetries in $X \in \{-1, 1\}^{2n}$. First symmetry to consider is $X = \{x_1, \ldots, x_{2n}\}$ and its negation $\overline{X} = \{\overline{x}_1 = -x_1, \ldots, \overline{x}_{2n} = -x_{2n}\}$.

**Lemma 2.2.** $\left(\prod X\right) F(n, X) = \left(\prod \overline{X}\right) F(n, \overline{X})$

*Proof.* First the product of the $X$'s is equal to the product of $\overline{X}$'s. There are two cases either $X$ contains an even number of $-1$'s, in which case $\overline{X}$ must contain an even number of $-1$'s. Or $X$ contains an odd number of $-1$'s, in which case $\overline{X}$ must contain an odd number of $-1$'s. They follow since $|X| = 2n$.

Next consider the polynomials $F(n, X)$ and $F(n, \overline{X})$,

$$F(n, X) = \prod_{i=1}^{n} \sum_{k=1}^{2n-i} x_k x_{k+i}$$

$$= \prod_{i=1}^{n} \sum_{k=1}^{2n-i} (-1)^2 x_k x_{k+i}$$

$$= \prod_{i=1}^{n} \sum_{k=1}^{2n-i} (-x_k)(-x_{k+i})$$

$$= F(n, \overline{X})$$

□

This symmetry halves the number of sequences of $X$ that need to be considered. A second symmetry to consider is $X = \{x_1, \ldots, x_{2n}\}$ and its reverse $\overleftarrow{X} = \{\overleftarrow{x}_1 = x_{2n}, \ldots, \overleftarrow{x}_{2n} = x_1\}$.

**Lemma 2.3.** $F(n, X) = F(n, \overleftarrow{X})$

*Proof.*

$$F(n, X) = \prod_{i=1}^{n} \sum_{k=1}^{2n-i} x_k x_{k+i}$$

$$= (x_1 x_2 + \ldots + x_{2n-1} x_{2n}) \ldots (x_1 x_{n+1} + \ldots + x_n x_{2n})$$

$$= (x_{2n} x_{2n-1} + \ldots + x_2 x_1) \ldots (x_{2n} x_n + \ldots + x_{n+1} x_1)$$

$$= (\overleftarrow{x}_1 \overleftarrow{x}_2 + \ldots + \overleftarrow{x}_{2n-1} \overleftarrow{x}_{2n}) \ldots (\overleftarrow{x}_1 \overleftarrow{x}_{n+1} + \ldots + \overleftarrow{x}_n \overleftarrow{x}_{2n})$$

$$= \prod_{i=1}^{n} \sum_{k=1}^{2n-i} \overleftarrow{x}_k \overleftarrow{x}_{k+i}$$

$$= F(n, \overleftarrow{X})$$

The third line follows since $\overleftarrow{x}_j = x_{2n-j+1}$. □

This symmetry almost halves the work, the exceptions are the palindromic sequence of $X$. A third kind of symmetry to consider is between $X = \{x_1, x_2, \ldots, x_{2n-1}, x_{2n}\}$ and the sequence where all the even positions are negated, $\widehat{X} = \{\widehat{x}_1 = x_1, \widehat{x}_2 = -x_2, \ldots, \widehat{x}_{2n-1} = x_{2n-1}, \widehat{x}_{2n} = -x_{2n}\}$.

**Lemma 2.4.** $\left(\prod X\right) F(n, X) = \left(\prod \widehat{X}\right) F(n, \widehat{X})$ when $n \equiv 0, 1 \pmod{4}$

*Proof.* First consider the case when $n \equiv 0 \pmod 4$.

In this case there are an even number of even position in the sequence. Among the even positions in $X$, either there are an even number of $-1$'s, in which case the even positions in $\widehat{X}$ must also contain an even number of $-1$'s. Or an odd number of the even positions in $X$ are $-1$, in which case an odd number of the even positions in $\widehat{X}$ must be $-1$. In both cases then the product of $X$ is equal to the product of $\widehat{X}$.

$$\prod X = \prod \widehat{X}$$

Next consider the polynomials $F(n, X)$ and $F(n, \widehat{X})$,

$$\begin{aligned}
F(n, X) &= \prod_{i=1}^{n} \sum_{k=1}^{2n-i} x_k x_{k+i} \\
&= \left( \prod_{odd(i)} \sum_{k=1}^{2n-i} x_k x_{k+i} \right) \left( \prod_{even(i)} \sum_{k=1}^{2n-i} x_k x_{k+i} \right) \\
&= \left( \prod_{odd(i)} \left( \sum_{odd(k)} x_k x_{k+i} + \sum_{even(k)} x_k x_{k+i} \right) \right) \left( \prod_{even(i)} \left( \sum_{odd(k)} x_k x_{k+i} + \sum_{even(k)} x_k x_{k+i} \right) \right) \\
&= \left( \prod_{odd(i)} \left( \sum_{odd(k)} (-1)^2 x_k x_{k+i} + \sum_{even(k)} (-1)^2 x_k x_{k+i} \right) \right) \left( \prod_{even(i)} \left( \sum_{odd(k)} x_k x_{k+i} + \sum_{even(k)} (-1)^2 x_k x_{k+i} \right) \right) \\
&= \left( \prod_{odd(i)} (-1) \left( \sum_{odd(k)} x_k(-x_{k+i}) + \sum_{even(k)} (-x_k) x_{k+i} \right) \right) \left( \prod_{even(i)} \left( \sum_{odd(k)} x_k x_{k+i} + \sum_{even(k)} (-x_k)(-x_{k+i}) \right) \right) \\
&= \left( \prod_{odd(i)} \left( \sum_{odd(k)} \widehat{x}_k \widehat{x}_{k+i} + \sum_{even(k)} \widehat{x}_k \widehat{x}_{k+i} \right) \right) \left( \prod_{even(i)} \left( \sum_{odd(k)} \widehat{x}_k \widehat{x}_{k+i} + \sum_{even(k)} \widehat{x}_k \widehat{x}_{k+i} \right) \right) \\
&= \prod_{i=1}^{n} \sum_{k=1}^{2n-i} \widehat{x}_k \widehat{x}_{k+i} \\
&= F(n, \widehat{X})
\end{aligned}$$

Note the $(-1)$ in the fifth line goes away since $n = 4m$ and there are an even number of odd numbers between 1 and $4m$.

$$\left( \prod X \right) F(n, X) = \left( \prod \widehat{X} \right) F(n, \widehat{X})$$

Next consider the case when $n \equiv 1 (\mod 4)$.

In this case there are an odd number of even positions in the sequence. Among the even positions in $X$, either there are an even number of $-1$'s, in which case the even positions in $\widehat{X}$ must contain an odd number of $-1$'s. Or an odd number of the even positions in $X$ are $-1$, in which case an even number of the even positions in $\widehat{X}$ must be $-1$. In both cases then the product of $X$ is negative of the product of $\widehat{X}$

$$\prod X = - \prod \widehat{X}$$

Now consider the evaluation of the polynomial. The proof is same as before except that the $(-1)$ in the fifth line stays since $n = 4m + 1$ and there are an odd number of odd numbers between 1 and $4m + 1$.

$$\begin{aligned}
F(n, X) &= \prod_{i=1}^{n} \sum_{k=1}^{2n-i} x_k x_{k+i} \\
&= (-1) \prod_{i=1}^{n} \sum_{k=1}^{2n-i} \widehat{x}_k \widehat{x}_{k+i} \\
&= -F(n, \widehat{X})
\end{aligned}$$

$$\left( \prod X \right) F(n, X) = \left( - \prod \widehat{X} \right) \left( -F(n, \widehat{X}) \right) = \left( \prod \widehat{X} \right) F(n, \widehat{X})$$

$\square$

This symmetry cuts the work in half, however it is only valid when a Skolem pairing exist, which is what we are interested in counting.

## 2.2 Results

This algebraic method is used for computing the number Skolem sequences when $n = 24, 25, 28, 29$ and the number of Langford sequences when $n = 27, 28$. We implemented the algorithm on NVIDIA's CUDA parallel computing platform. The first issue is parallelization. Since each sum can be computed independently, we can easily split $\{1, -1\}^{2n}$ sequences into several independent sets and simply put the result together at the end. To avoid rounding errors we only allow splitting into log 2 divisible numbers. This splitting is done in two levels, first across multiple GPU's, and second with in each GPU. We note that the values for $x_{2n}$ and $x_{2n-1}$ can be fix due to the first and third symmetries.

$$\underbrace{x_{2n} x_{2n-1}}_{symmetry} \underbrace{x_{2n-2} \ldots x_{47}}_{GPU's} \underbrace{x_{46} \ldots x_{27}}_{threads} x_{26} \ldots x_1$$

Each GPU kernel is configured to run $2^{20}$ threads where each thread will compute $2^{26}$ sums. With this geometry when $n = 24$, $S(24)$ is run on a single card with a running time of less than a day. When $n = 25$, $S(25)$ is split into four GPU jobs, and submitted to the HPCC cluster, all jobs completed in less than a day. When $n = 27$, $L(27)$ is split into 64 GPU jobs, and submitted to the HPCC cluster, where 32 jobs were run simultaneously, all jobs completed in about two days. When $n = 28$, the computations where split into 256 GPU jobs, submitted to the HPCC cluster, with 32 simultaneously running jobs. The counts for $S(28)$ and $L(28)$ each completed in about nine days. When $n = 29$, the computations where split into 1024 GPU jobs, submitted to the HPCC cluster. All jobs completed in about eight weeks. The second issue is dealing with "large" numbers. This problem is handled by carrying out the calculations modulo number of co-primes, then the actual value is obtained using the chinese remainder theorem. The results are summarized in tables 4 and 5.

## 3 Approximate Counting Algorithm

The best known exact algorithms for counting the number of Skolem sequences take exponential time. We can relax the exactness requirement and compute an approximate count in polynomial time. The general approach is to compute the ratio between the number of solutions of the original problem and a simplified version of the problem whose exact count is easy to find. Then the count is simply the product of this ratio with the count of the simplified problem. However, computing the exact ratio is as difficult as counting exactly. Therefore, the ratio is approximated using Monte Carlo simulation scheme in the span of the configuration space of all solutions to the simplified problem. A common issue with increasing problem size is the ratio tends to zero. This occurs when the numbers of solutions of the original and simplified version of the problem differ by many orders of magnitudes. To compensate, a set of intermediate problems are introduced and the ratio is seen as the product of the ratios of the intermediate problems.

$$\frac{|O|}{|S|} = \frac{|S_1|}{|S|} \times \frac{|S_2|}{|S_1|} \times \frac{|S_3|}{|S_2|} \times \ldots \times \frac{|O|}{|S_k|}$$

Consider the simple problem of placing $n$ pairs of colored balls into $2n$ bins, where each pair of $k$ colored ball are at distance $k$ from each other. It is easy to see there are $\frac{(2n-1)!}{(n-1)!}$ ways to fill the bins such that color $k$ appears in bins $i$ and $i+k$. The Skolem sequences are the solutions to the simplified problem were all the bins are occupied. To create intermediate problems, the configurations

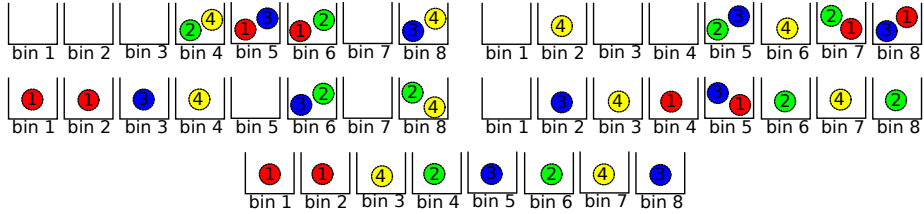

Figure 1: Example configurations when $n = 4$, with energies 0 to 4

are associated with a statistical system whose partition function is

$$Z(\beta) = \sum_C \exp\left[-\beta E(C)\right]$$

where $\beta$ is the reciprocal of the temperature, and the energy of the system, $E(C)$, is the number of empty bins. Each intermediate problem, $S_i$, is identified with reciprocal temperature $\beta_i$. At infinite temperature, $\beta = 0$, the function counts the solutions to the simplified problem, that is $Z(0) = \frac{(2n-1)!}{(n-1)!}$. At zero temperature, $\beta = \infty$, when all bins are occupied, the energy of the system is at a minimum, $E(C) = 0$, and $Z(\beta)$ counts the number Skolem sequences. Let $0 = \beta_1 < \beta_2 < \ldots < \beta_m < \beta$, then

$$\frac{Z(\beta)}{Z(0)} = \frac{Z(\beta_2)}{Z(\beta_1)} \frac{Z(\beta_3)}{Z(\beta_2)} \cdots \frac{Z(\beta)}{Z(\beta_m)}$$

The expected values of the ratios are approximated using a Monte Carlo [10] procedure. The procedure relies on repeated random sampling. The samples are produced iteratively, with the distribution of the next sample being dependent only on the current sample. Specifically, starting with a random placement of balls into bins, at each iteration the algorithm picks a candidate for the next sample by uniformly at random choosing a color $k$, and moving pair of $k$-colored balls uniformly at random to a new position. And accepting the candidate with probability,

$$p = \min\left[1, e^{-\beta \Delta E}\right]$$

and continuing with the candidate as the new sample otherwise rejecting it and continuing with old sample.

For large values of $\beta$, the acceptance rates becomes rather small and the simulations could run into problems. The situation is improved by using the Parallel Tempering [10]. This technique allows the system at high temperature to feed new configurations to local optimizer of a system at low temperature. In effect allowing tunneling between meta stable states to improve convergence to a global optimum. To achieve this tunneling, in addition to updating independently the configurations, neighboring configurations of adjacent $\beta_i$-values are exchanged with probability

$$p_{i,i+1} = \min\left[1, e^{-(\beta_{i+1} - \beta i)(E(C_i) - E(C_{i+1}))}\right]$$

```
Algorithm 2: Approximate Counting
  Data: N number of pairs, β inverse temperatures, M ← |β|
  Result: S approximate count
  for i ← 1 to M do
   │ C(i) ← a random configuration;
  end
  for i ← 1 to iterations do
     for j ← 1 to M do
      │ Ratio(j) ← Ratio(j) + (e^{-(β(j+1)-β(j))*E(C(j))});
     end
     for j ← 1 to M do
        N ← Random Neighbor(C(j));
        p ← min [1, e^{-(E(N)-E(C(j)))β(j)}];
        if p ≤ random(0,1) then
         │ C(j) ← N;
        end
     end
     for j ← 1 to M do
        p ← min [1, e^{-(β(j+1)-β(j))(E(C(j))-E(C(j+1)))}];
        if p ≤ random(0,1) then
         │ C(j), C(j+1) ← C(j+1), C(j);
        end
     end
  end
  S ← (2n-1) × ... × n;
  for i ← 1 to M do
   │ S ← S × Ratio(i)/iterations;
  end
```

In order to make the procedure efficient, the inverse temperatures are chosen such that the acceptance rates for the configuration exchanges are no smaller than one half. Furthermore, number of temperatures are limited as having too many systems hampers the rapid exchange of information from higher to lower temperatures and vice versa.

## 3.1 Results

The procedure is implemented in $c++$ standard version 11, on a system running Fedora release version twenty four, compiled with GNU compiler version 6.2. A sample run for $n = 12$ is shown in table 6, with twelve temperature levels and $2^{24}$ iterations. The approximation results for the Skolem and Langford sequences are summarized in tables 7 and 8 respectively. We run the procedure for each $n$ several times and take average value of the runs as the approximate value. Although we proposed the algorithm without any proof, experimentally we see that the error found is less than one percent.

| $i$ | $\beta$ | $W1$ | $W2$ | $E[e]$ | $E[z_{i+1}/z_i]$ |
|---|---|---|---|---|---|
| 1 | 0 | 1 | 0.57 | 8.60 | 0.013241 |
| 2 | 0.54 | 0.82 | 0.56 | 7.40 | 0.0221532 |
| 3 | 1.1 | 0.64 | 0.56 | 6.21 | 0.0363131 |
| 4 | 1.69 | 0.48 | 0.56 | 5.04 | 0.056943 |
| 5 | 2.33 | 0.34 | 0.57 | 3.93 | 0.100785 |
| 6 | 3 | 0.23 | 0.57 | 2.94 | 0.164186 |
| 7 | 3.73 | 0.15 | 0.52 | 2.03 | 0.241043 |
| 8 | 4.65 | 0.098 | 0.55 | 1.10 | 0.437722 |
| 9 | 5.82 | 0.069 | 0.70 | 0.40 | 0.691262 |
| 10 | 8.1 | 0.06 | 0.95 | 0.04 | 0.957744 |
| 11 | 16 | 0.059 | 0.99 | $8.89 \times 10^{-05}$ | 0.999933 |
| 12 | 32 | 0.059 | 0.99 | $4.43 \times 10^{-05}$ | 0.99997 |

Table 6: A sample run to approximate $S(12)$, $W1$ is the acceptance of new configuration, $W2$ is the exchange rate of neighboring configurations, the approximate value then is $S(12) \approx \left(\frac{1}{2}\right)\left(\frac{23!}{11!}\right)\prod E[z_{i+1}/z_i] = 227009$

| $n$ | exact count | approximate count | error |
|---|---|---|---|
| 4 | 3 | 3 | $\sim 0\%$ |
| 5 | 5 | 5 | $\sim 0\%$ |
| 8 | 252 | 252 | $\sim 0\%$ |
| 9 | 1,328 | 1,328 | $\sim 0\%$ |
| 12 | 227,968 | $2.265 \times 10^5$ | $\sim -0.66\%$ |
| 13 | 1,520,280 | $1.520 \times 10^6$ | $\sim -0.01\%$ |
| 16 | 700,078,384 | $7.009 \times 10^8$ | $\sim +0.12\%$ |
| 17 | 6,124,491,248 | $6.139 \times 10^9$ | $\sim +0.24\%$ |
| 20 | 5,717,789,399,488 | $5.733 \times 10^{12}$ | $\sim +0.27\%$ |
| 21 | 61,782,464,083,584 | $6.186 \times 10^{13}$ | $\sim +0.12\%$ |
| 24 | 102,388,058,845,620,672 | $1.026 \times 10^{17}$ | $\sim +0.19\%$ |
| 25 | 1,317,281,759,888,482,688 | $1.317 \times 10^{18}$ | $\sim -0.03\%$ |
| 28 | 3,532,373,626,038,214,732,032 | $3.523 \times 10^{21}$ | $\sim -0.27\%$ |
| 29 | 52,717,585,747,603,598,276,736 | $5.294 \times 10^{22}$ | $\sim +0.42\%$ |
| 32 | | $2.213 \times 10^{26}$ | |
| 33 | | $3.614 \times 10^{27}$ | |

Table 7: Number of Skolem Sequences, oeis $A059106$, the approximate value found using algorithm 2

## 4 Conclusion

Godfrey's algebraic approach to count the number of sequences reduced the computation time for counting sequences by a factor of $n^n$. This breakthrough enabled researchers to count the number of sequences beyond $n = 17$. With the massive parallelism available in the Graphics Processing Units we were able to extend the count further and obtain the count for all values less than thirty. With ever improving hardware it may become possible to count $S(32)$ and $S(33)$, however it is unlikely that exact values could be found using the algebraic method for any values beyond that. Certainly it will be impossible to count $S(64)$, unless there is another breakthrough.

| $n$ | exact count | approximate count | error |
|---|---|---|---|
| 3 | 1 | 1 | $\sim 0\%$ |
| 4 | 1 | 1 | $\sim 0\%$ |
| 7 | 26 | 26 | $\sim 0\%$ |
| 8 | 150 | 150 | $\sim 0\%$ |
| 11 | 17,792 | $1.779 \times 10^4$ | $\sim 0\%$ |
| 12 | 108,144 | $1.087 \times 10^5$ | $\sim +0.5\%$ |
| 15 | 39,809,640 | $3.976 \times 10^7$ | $\sim -0.1\%$ |
| 16 | 326,721,800 | $3.267 \times 10^8$ | $\sim -0.007\%$ |
| 19 | 256,814,891,280 | $2.558 \times 10^{11}$ | $\sim -0.38\%$ |
| 20 | 2,636,337,861,200 | $2.621 \times 10^{12}$ | $\sim -0.57\%$ |
| 23 | 3,799,455,942,515,488 | $3.781 \times 10^{15}$ | $\sim -0.48\%$ |
| 24 | 46,845,158,056,515,936 | $4.649 \times 10^{16}$ | $\sim -0.7\%$ |
| 27 | <span style="color:red">111,683,611,098,764,903,232</span> | $1.115 \times 10^{20}$ | $\sim -0.1\%$ |
| 28 | <span style="color:red">1,607,383,260,609,382,393,152</span> | $1.603 \times 10^{21}$ | $\sim -0.3\%$ |
| 31 | | $5.381 \times 10^{24}$ | |
| 32 | | $8.812 \times 10^{25}$ | |

Table 8: Number of Langford Sequences, oeis $A014552$, the approximate value found using algorithm 2

| $n$ | Backtracking<br>$1-CPU$ | Algebraic Algorithm | | |
|---|---|---|---|---|
| | | $1-CPU$ | $1-GPU$ | $multi-GPU$ |
| 12 | $1.4 sec$ | $0.3 sec$ | | |
| 13 | $12 sec$ | $1.2 sec$ | | |
| 16 | $175.25 min$ | $1.38 min$ | $0.5 sec$ | |
| 17 | $\sim 1.28 days$ | $6.18 min$ | $2.5 sec$ | |
| 20 | | $\sim 7.18 hrs$ | $4.08 min$ | |
| 21 | | $\sim 1.24 days$ | $17.67 min$ | |
| 24 | | | $21.55 hrs$ | |
| 25 | | | | $\sim 1 day$ |
| 28 | | | | $\sim 1.28 wks$ |
| 29 | | | | $\sim 8 wks$ |

Table 9: Running times of the backtracking algorithm and Godfrey's algebraic algorithm for computing number of Skolem sequences on $i7-3770$ intel CPU and NVIDIA Kepler GPU. Backtracking algorithm's running time $O(4^n n^n)$. Godfrey's algebraic algorithm's running time $O(4^n)$.

In section 3 we proposed an approximation procedure and experimentally observed that the approximate values found were within one percent of the exact values. We can even find an approximate value for $S(64) \approx 4.1 \times 10^{70}$. However we do not know or can prove the quality of this value.

**Acknowledgment:** We like to thank the CUNY HPCC for the use of their computational cluster. The CUNY HPCC is operated by the College of Staten Island and funded, in part, by grants from the City of New York, State of New York, CUNY Research Foundation, and National Science Foundation Grants CNS-0958379, CNS-0855217 and ACI 1126113.